\documentclass[aps,prl,twocolumn,superscriptaddress,showpacs]{revtex4}
\usepackage[usenames]{color}
\newcommand{\comment}[1]{}
\usepackage{graphicx}
\usepackage{amsmath}
\usepackage{epsfig}
\usepackage{eufrak}

\newcommand{\comm}[2]{\left[ {#1},{#2} \right]}
\newcommand{\cn}[2]{c_{{#1},{#2}}}
\newcommand{\cy}[2]{c^\dagger_{{#1},{#2}}}
\newcommand{\dn}[2]{d_{{#1}+{\bf Q},{#2}}}
\newcommand{\dy}[2]{d^\dagger_{{#1}+{\bf Q},{#2}}}

\def\be{\begin{eqnarray}}
\def\ee{\end{eqnarray}}
\def\nn{\nonumber}
\def\l{\left}
\def\rr{\right}

\def\lofa{LaOFeAs}
\def\loffa{{LaO$_{1-x}$F$_x$FeAs}}

\def\coffa{CeO$_{1-x}$F$_x$FeAs}

\def\bQ{{\bf Q}}
\def\bk{{\bf k}}
\def\half{\frac{1}{2}}
\def\up{\uparrow}
\def\down{\downarrow}

\begin{document}

\title{Collective Modes and Emergent SO(6) Symmetry in the Iron Pnictides}
\date{\today}
\author{Daniel Podolsky}
\affiliation{Department of Physics, University of Toronto,
Toronto, Ontario M5S 1A7, Canada}
\affiliation{Physics Department, Technion, Haifa 32000, Israel}
\author{Hae-Young Kee}
\affiliation{Department of Physics, University of Toronto,
Toronto, Ontario M5S 1A7, Canada}
\author{Yong Baek Kim}
\affiliation{Department of Physics, University of Toronto,
Toronto, Ontario M5S 1A7, Canada}

\begin{abstract}
We show the existence of an emergent SO(6) symmetry in the low energy description of the iron pnictides.  This approximate symmetry provides a unifying framework for the occurrence of spin density wave (SDW) and superconductivity (SC) in these materials.  We use this symmetry to make several predictions for future experiments, including the topology of the phase diagram and the presence of various resonant modes in neutron scattering experiments in both the SC and SDW phases.  We also predict the existence of a new ``Orbital Density Wave'' state, which competes with both SDW and SC orders.
\end{abstract}

\pacs{74.20.De, 74.25.Ha}

\maketitle

{\em Introduction.---}
The recent discovery of superconductivity in LaOFeP \cite{kamihara06} and subsequently in \loffa\ \cite{kamihara08} has generated great excitement in the iron pnictides, a new class of materials with great potential for applications, with superconducting $T_c$ above 50 K already been achieved \cite{xhchen,gfchen}.  The phase diagram in these materials displays both spin density wave (SDW) with
the ordering wavevector ${\bf Q}=(\pi,0)$ and superconducting (SC) orders \cite{luetkens,zhao}.  For instance, \lofa\ is a metal with SDW order, which undergoes a first order transition to a superconductor upon doping \cite{luetkens}.  Similarly, \coffa\ undergoes a doping-tuned continuous transition from the SDW to a SC \cite{zhao}.

In this Letter, we investigate the interplay between these two types of order in terms of 
symmetry considerations.  It has been suggested that a strong contender for the SC order parameter is an ``odd-sign" $s$-wave where the SC order parameters on the electron and hole pockets acquire opposite signs while
there is a full gap at both pockets \cite{mazin,chubukov,tesanovic,wang}. 
Here, we will show that there is a broad class of Hamiltonians with SO(6) symmetry that naturally give rise to both the ``odd-sign" SC order state and to an SDW with ordering wave vector ${\bf Q}=(\pi,0)$.
Moreover, the symmetry analysis reveals a hidden order parameter, dubbed ``Orbital Density Wave" (ODW).
This ODW is intimately connected to the SDW and SC orders via the SO(6) group; namely they form a six dimensional 
vector under the action of SO(6). The SO(6) symmetry is emergent in the sense that it becomes an increasingly better symmetry in the limit of longer distances and lower energies. We explore experimental consequences of such an emergent symmetry in iron pnictides.

\begin{figure}
\includegraphics[width=2.5in]{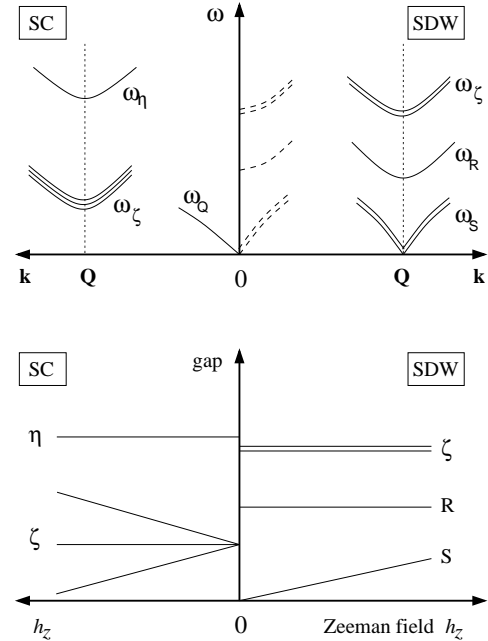}
\caption{Top panel: Collective excitations in the SC and SDW phases.  Inside the SC, there are degenerate gapped modes $\zeta_{x,y,z}$, and a higher energy $\eta$ mode.  The $Q$ mode, shown here as a linearly dispersing Goldstone mode, acquires a gap when long-range Coulomb repulsion is taken into account.  In the SDW state (say, $\Phi_x$), there are two linearly dispersing magnon modes $S_y$ and $S_z$, and gapped $R_x$ and $\zeta_x^{\dagger}$ modes.  The relative size of the $R_x$ and $\zeta_x$ gaps may depend on doping.  Lower panel: Gap under an applied Zeeman field for different collective modes.   In the SC, the $\zeta_{x,y,z}$ modes are split, as shown.  Inside the SDW phase, the magnon mode $S_y$ obtains a gap.  All other modes are unaffected to linear order in field.
\label{fig:collective} } \vskip-0.3in
\end{figure}

The idea of enhanced symmetry in strongly correlated electron systems was first demonstrated in the Hubbard model at half-filling \cite{yang89}, and it has been used to describe various systems such as the cuprates \cite{Zhang97,demlerRMP,kee07} and the organic superconductors \cite{Murakami2000,podolsky04}.  The iron pnictides may be particularly well-suited for such an approach.   
The Coulomb interaction in these materials may be relatively small, as seen in the metallic nature of the magnetic state.  This is in contrast to the cuprates, where the interplay of SC and magnetism is complicated by Mott physics.

The emergent SO(6) symmetry leads to several predictions for the phase diagram and excitation spectrum.  The resulting excitation spectra are summarized in Fig.\ref{fig:collective}. For instance, inside the SC state, we predict the existence of three low energy $S=1$ excitations (denoted $\zeta$) with orbital angular momentum $L=0$, and charge $q=2$ -- analogous to the $\pi$-excitation proposed in the context of the cuprates \cite{demlerRMP} -- 
and a low energy $S=0$ excitation (denoted $\eta$), which carries $L=1$, and $q=2$.  
These modes can be probed by inelastic neutron scattering experiments via 
$S=1$ or $L=1$ quantum numbers.
On the other hand, in the SDW state, we predict a low energy $S=1$ (denoted $R$) excitation
with $L=1$, and $q=0$, which couples to neutrons, and two other excitations (denoted $\zeta$), 
with $S=1$, $L=0$, and $q=2$, which do not couple to neutrons.

{\em Effective Hamiltonian.---}
The band structure of the iron pnictides involves 
all five $d$-orbitals of the Fe atoms.  Staggering of the out-of plane position of the As atoms leads to doubling of the unit cell to include two Fe atoms. 
The band structure has been computed within the local density approximation \cite{singh} and confirmed experimentally in ARPES \cite{ding} and quantum oscillation \cite{sebastian} experiments.  The Fermi surface is composed of two electron pockets surrounding the point $(0,0)$ of the Brillouin zone, and two hole pockets around $\bQ=(\pi,0)$ (or, equivalently, $(0,\pi)$).  The electron and hole pockets are similar in size, leading to an approximate particle-hole symmetry.  In this analysis we concentrate on one electron pocket and one hole pocket, since the physics governing ordering instabilities is already contained at this level (although the other bands are necessary to understand the precise ordering pattern of the SDW and ODW orders within one unit cell).

In a two-pocket model, the most general on-site interaction Hamiltonian is,
\begin{eqnarray}
	\label{eq:genInt}
H'= \frac{1}{2}\sum_{i\sigma\sigma'} \sum_{\alpha\beta\gamma\delta=c,d}U^{\alpha\beta}_{\gamma\delta}\psi^\dagger_{i\alpha\sigma}\psi^\dagger_{i\beta\sigma'} \psi_{i\delta\sigma'} \psi_{i\gamma\sigma},
\end{eqnarray}
where, $\psi^\dagger_{i\alpha\sigma}$ creates an electron with spin $\sigma$ at site $i$.  $\alpha\cdots\delta$ are pocket indexes: $\psi^\dagger_{ic\sigma}= c^\dagger_{i\sigma}$ and $\psi^\dagger_{id\sigma}= d^\dagger_{i\sigma}$ create electrons near the electron  and hole pockets, respectively.  There are five independent coupling constants, $u_1\equiv U^{cd}_{cd}=U^{dc}_{dc}$, $u_2\equiv U^{cd}_{dc}=U^{dc}_{cd}$, $u_3\equiv U^{cc}_{dd}=(U^{dd}_{cc})^*$, $u_4\equiv U^{cc}_{cc}$, and $u_5\equiv U^{dd}_{dd}$;  all other interactions are zero.  For systems with particle-hole symmetry, $u_4=u_5$.  The Hamiltonian (\ref{eq:genInt}) possesses SO(6) symmetry provided that the following conditions are satisfied:
\begin{eqnarray}
\label{eq:effInt}
u_2= 0\ \ \ {\rm and}\ \ \  u_4=-u_1,
\end{eqnarray}
and $u_3$ can take any value.  This defines a 2 dimensional subspace of SO(6) symmetric Hamiltonians.  Remarkably, a one-loop RG analysis of the two band Hubbard model with particle-hole symmetry yields an RG that flows towards a fixed point satisfying the SO(6) conditions (\ref{eq:effInt}) \cite{chubukov}.  The RG is controlled by the size of the fermi pockets, and in the limit of infinitesimally small pockets the RG flows asymptotically to an SO(6) symmetric point.


{\em The group SO(6).---}
One can show that the interaction Hamiltonian at with the constraints (\ref{eq:effInt})
commutes with the following 15 generators of SO(6),
\begin{eqnarray}
\label{eq:generators}
\hat{Q} &=&-\half\sum_{\bk\sigma}\l( \cy{\bk}{\sigma}\cn{\bk}{\sigma} +\dy{\bk}{\sigma}\dn{\bk}{\sigma}-1\rr) \cr
\hat{S}_\alpha &=&\half\sum_{\bk\sigma\sigma'}\l( \cy{\bk}{\sigma}\sigma^\alpha_{\sigma\sigma'}\cn{\bk}{\sigma'} +\dy{\bk}{\sigma}\sigma^\alpha_{\sigma\sigma'}\dn{\bk}{\sigma'}\rr) \cr
\hat{R}_\alpha &=&\half\sum_{\bk\sigma\sigma'}\l( \cy{\bk}{\sigma}\sigma^\alpha_{\sigma\sigma'}\cn{\bk}{\sigma'} -\dy{\bk}{\sigma}\sigma^\alpha_{\sigma\sigma'}\dn{\bk}{\sigma'}\rr) \cr
\hat{\zeta}^+_\alpha &=&\sum_{\bk\sigma\sigma'} \cy{\bk}{\sigma}\l(\hat{\sigma}_\alpha\hat{\sigma}_y\rr)_{\sigma\sigma'}\dy{-\bk}{\sigma'},\ \ \ \hat{\zeta}^-=(\hat{\zeta}^+)^\dagger \cr
\hat{\eta}^+ &=&-i\sum_{\bk\sigma\sigma'} \cy{\bk}{\sigma}\sigma^y_{\sigma\sigma'}\dy{-\bk}{\sigma'},\ \ \ \ \ \ \hat{\eta}^-=(\hat{\eta}^+)^\dagger
\end{eqnarray}
where $\alpha$ takes the values $x$, $y$, $z$, and 
${\hat \sigma}_{\alpha}$ are Pauli matrices.
The 15 generators are split into 5 groups: ({\em i}) the charge operator $\hat{Q}$; ({\em ii}) three spin operators $\hat{S}_\alpha$; ({\em iii}) three operators $\hat{R}_\alpha$, which measure the difference in spin between pockets $c$ and $d$; ({\em iv}) the six operators $\hat{\zeta}_\alpha^\pm$; and ({\em v}) the two operators $\hat{\eta}^\pm$.  Both $\hat{\zeta}_\alpha^+$ and $\hat{\eta}^+$ create pairs of electrons, $[\hat{Q},\hat{\zeta}^+_\alpha]=-\hat{\zeta}^+_\alpha,$ and $[\hat{Q},\hat{\eta}^+]=-\hat{\eta}^+,$ but they differ in that $\hat{\eta}^+$ creates a spin-singlet pair, whereas $\hat{\zeta}_\alpha^+$ creates a spin-triplet pair with zero spin projection along the $\alpha$ direction.  This difference is reflected in their commutator with $S_\alpha$: $[\hat{S}_\alpha,\hat{\zeta}_\beta^+]= i\epsilon_{\alpha\beta\gamma}\hat{\zeta}_\gamma^+$ and $[\hat{S}_\alpha,\hat{\eta}^+]=0$.

We can organize the 15 generators (\ref{eq:generators}) by collecting them into an antisymmetric 6$\times$6 matrix of operators, 
\begin{eqnarray}
\hat{L}_{ab}=\left(\begin{array}{c c c c c c}
0 & \hat{Q} & \Re\,\hat{\zeta}_x & \Re\,\hat{\zeta}_y & \Re\,\hat{\zeta}_z & \Re\,\hat{\eta} \\
\, & 0 & \Im\,\hat{\zeta}_x & \Im\,\hat{\zeta}_y & \Im\,\hat{\zeta}_z & \Im\,\hat{\eta} \\
\, & \, & 0 & \hat{S}_z & -\hat{S}_y & \hat{R}_x \\
\, & \, & \, & 0 & \hat{S}_x &  \hat{R}_y \\
\, & \, & \, & \, & 0 & \hat{R}_z \\
\, & \, & \, & \, & \, & 0
\end{array}
\right)
\label{eq:genMatrix}
\end{eqnarray}
where $\Re\,\hat{\cal O}\equiv\half(\hat{\cal O}^-+\hat{\cal O}^+)$ and $\Im\,\hat{\cal O}\equiv\frac{1}{2i}(\hat{\cal O}^--\hat{\cal O}^+)$, 
and the missing entries are obtained via $\hat{L}_{ab}=-\hat{L}_{ba}$.   In the matrix form (\ref{eq:genMatrix}), we can state the commutation relations between all 15 operators succinctly:
\begin{eqnarray}
\label{eq:LLComm}
\comm{\hat{L}_{ab}}{\hat{L}_{cd}}=-i\l(\delta_{ad}\hat{L}_{bc}+\delta_{bc}\hat{L}_{ad}-\delta_{bd}\hat{L}_{ac}-\delta_{ac}\hat{L}_{bd}\rr).\nn
\end{eqnarray}
This is the Lie algebra of the group SO(6).

{\em SO(6) as a symmetry of the effective Hamiltonian.---}  In order to show that SO(6) is an emergent symmetry of the iron pnictides, we now consider the commutators of $\hat{L}_{ab}$ with the full Hamiltonian, $H=H_0+H'$, where
\begin{eqnarray}
H_0=\sum_{\bk\sigma}\l(\epsilon_{\bk}^{c} \cy{\bk}{\sigma}\cn{\bk}{\sigma}+\epsilon_{\bk}^{d} \dy{\bk}{\sigma}\dn{\bk}{\sigma} \rr)
\end{eqnarray}
is the free Hamiltonian. Note that $H_0$ conserves total charge $Q$, total spin $S_\alpha$, and the relative spin $R_\alpha$.  Hence, the only non-trivial commutators of $H_0$ are,
\begin{eqnarray}
\comm{H_0}{\zeta_\alpha^+}&=&\sum_{\bk\sigma\sigma'}\l(\epsilon_{\bk}^{c}+\epsilon_{-\bk}^{d}\rr)\cy{\bk}{\sigma}\l(\hat{\sigma}_\alpha\hat{\sigma}_y\rr)_{\sigma\sigma'}\dy{-\bk}{\sigma'},\cr
\comm{H_0}{\eta^+}&=&-i\sum_{\bk\sigma\sigma'}\l(\epsilon_{\bk}^{c}+\epsilon_{-\bk}^{d}\rr)\cy{\bk}{\sigma}\sigma^y_{\sigma\sigma'}\dy{-\bk}{\sigma'}.
\end{eqnarray}
In the iron pnictides, the electron and hole pockets have similar shapes and sizes \cite{palee}.  We can then tune the chemical potential to a point where the system has an approximate particle-hole symmetry, with the low energy dispersion $\epsilon_{\bk}^c+\epsilon_{-\bk}^d\approx 0$ (nesting condition).  At this point the free Hamiltonian $H_0$ has an approximate SO(6) symmetry. Since $H'$ commutes with the full SO(6) generators, $H = H_0+H'$ has an approximate SO(6) symmetry.

In practice, there are three sources of SO(6) symmetry breaking.  First, the particle and hole pockets do not have identical shapes, so that the particle-hole symmetry is only approximate.  Second, doping the system away from the (approximate) particle-hole symmetric point gives,
\begin{equation}
\comm{H_0}{\hat{\zeta}^+_\alpha} \approx 2 {\bar \mu} \, \hat{\zeta}^+_\alpha, \hskip 0.5cm 
\comm{H_0}{\hat{\eta}^+} \approx 2 {\bar \mu} \, \hat{\eta}^+,
\end{equation}
where ${\bar \mu}$ is the shift of the chemical potential relative to the particle-hole symmetric point.  In this case, the symmetry of $H_0$ is reduced to SO(4)$_{SR}\times$U(1)$_Q$, the group generated by $\hat{S}_\alpha$, $\hat{R}_\alpha$, and $\hat{Q}$.  Increased doping reduces the nesting between $c$ and $d$ pockets, and therefore suppresses SDW order in favor of SC.  By changing doping, we can tune to the SO(6) symmetric point.  Finally, there are likely to be residual interactions that do not satisfy Eq. (\ref{eq:effInt}).  However, provided that these are not strong, the $\zeta$, $\eta$, and $R$ modes will remain sharp excitations, and the topology of the phase diagram will be unaffected.

{\em Ground state manifold.---}
Six different order parameters are connected by SO(6): superconductivity ($\hat{\Delta}^\pm$), SDW order ($\hat{\Phi}_x$, $\hat{\Phi}_y$, and $\hat{\Phi}_z$), and ``Orbital Density Wave'' (ODW) order ($\hat{\Xi}$).   For $u_3>0$ interactions favor the formation of ``odd-sign'' $s$-wave superconductivity, with a relative minus sign
between the electron and hole pockets,
\begin{eqnarray}
\hat{\Delta}^+=\sum_{\bf k}\l(\cy{\bk}{\up}\cy{-\bk}{\down}-\dy{\bk}{\up}\dy{-\bk}{\down}\rr),\ \ \hat{\Delta}^-=(\hat{\Delta}^+)^\dagger
\end{eqnarray}
$u_3>0$ also favors a SDW instability with momentum $\bQ$,
\begin{eqnarray}
\hat{\Phi}_\alpha=\half\sum_{\bk\sigma\sigma'}\l(\cy{\bk}{\sigma}\sigma^\alpha_{\sigma\sigma'}\dn{\bk}{\sigma'}+\dy{\bk}{\sigma}\sigma^\alpha_{\sigma\sigma'}\cn{\bk}{\sigma}\rr).
\end{eqnarray}
Finally, as we will show, it also favors the ODW,
\begin{eqnarray}
\hat{\Xi}=-\frac{i}{2}\sum_{\bk\sigma}\l(\cy{\bk}{\sigma}\dn{\bk}{\sigma}-\dy{\bk}{\sigma}\cn{\bk}{\sigma}\rr).
\end{eqnarray}
In real space, $\Xi=-\frac{i}{2}\sum_{r} (-1)^{r_x}(n_{r,c+id}-n_{r,c-id})$, where $n_{r,c\pm id}=\frac{1}{2}\sum_\sigma(c_{r\sigma}\pm id_{r\sigma})^\dagger(c_{r\sigma}\pm id_{r\sigma})$.  Thus, the ODW is a state with alternating $c+id$ and $c-id$ orbitals along the $x$ direction -- it breaks time reversal symmetry and translational symmetry by one lattice spacing along $x$, but preserves the combination of the two.  It also preserves spin rotational symmetry.

These six order parameters can be organized into a vector $\hat{n}_a$, $a=1\ldots 6$, where
$\hat{n}_1=\Re\,\hat{\Delta}$, $\hat{n}_2=\Im\,\hat{\Delta}$,
$\hat{n}_3=\hat{\Phi}_x$,
$\hat{n}_4=\hat{\Phi}_y$,
$\hat{n}_5=\hat{\Phi}_z$, and
$\hat{n}_6=\hat{\Xi}$.
Then, we find 
\begin{eqnarray}
\comm{\hat{L}_{ab}}{\hat{n}_{c}}=-i \l(\delta_{bc}\hat{n}_a+\delta_{ac}\hat{n}_b\rr).
\label{eq:LnComm}
\end{eqnarray}
Hence, the ground state order parameters transform as a six dimensional vector under the action of SO(6).

{\em Phase diagram.---}
As a first application of SO(6) symmetry, we consider the phase diagram of the iron pnictides.  Symmetry places strong constraints on the Ginsburg-Landau (GL) free energy, which in turn constraints the possible topologies of the phase diagram.  To quartic order in the order parameters, the most general GL free energy is, 
\begin{eqnarray}
F_{GL} &=& \int d^3{\bf r} \l[ \half(\nabla {\bf n})^2+ r {\bf n}^2 + u \l({\bf n}^2 \rr)^2 \rr. \nn\\
&\,&-r_d(n_1^2+n_2^2-n_3^2-n_4^2-n_5^2-n_6^2)\nn\\
&\,&\l. -r_H(n_3^2+n_4^2+n_5^2-3 n_6^2)\rr] + F'_Q. \label{eq:GL}
\end{eqnarray}
Here, the first line is SO(6)-symmetric, while all other terms break the symmetry explicitly. The sign and magnitude of $r_d$ can be tuned experimentally through doping.  The term $F'_Q$ -- not written explicitly -- contains all possible quartic terms that break SO(6) symmetry, and is expected to be small.

We assume that $r_H$ is small in magnitude, and first consider the case $r_H=0$.  For simplicity, we will also assume that $F'_Q$ is SO(4)$_{RS}\times$U(1)$_Q$ symmetric.  A renormalization group analysis of such a free energy was carried out in Ref. \onlinecite{Murakami2000}.  There, the phase diagram was found to have three possible topologies, depending on the coefficient $g$ appearing in $F_Q'=g (|\Delta|^2-{\bf \Phi}^2-\Xi^2)^2+\ldots$.  For $g>0$, there is a tetracritical point where four phases meet: the normal state, SC, magnetism, and a mixed state of coexisting magnetism and SC.  For $g<0$, there is a direct first order transition between SC and magnetism, and first order lines ending at tricritical points separating the ordered and normal states.  For the special case of full SO(6) symmetry, $g=0$, there is a first order transition between SC and magnetism ending at a bicritical point, with second order lines separating the ordered and normal states.

The precise nature of the magnetic state appearing in the phase diagram depends on the sign of $r_H$.  For $r_H>0$, the magnetic state will be the SDW, whereas for $r_H<0$, it will be the ODW. Since a ferromagnetic Hund's rule coupling gives a positive contribution to $r_H$, we concentrate on the SDW in what follows.

{\em Low energy excitations.---}
In order to study the low energy excitations that arise due to the approximate SO(6) symmetry, we introduce a quantum rotor model \cite{demlerRMP},
\begin{eqnarray}
H_{\rm QR}=\frac{1}{2\chi}\sum_{i,a<b} \hat{L}_{i,ab}^2-\sum_{\langle ij\rangle,a} r_a \hat{n}_i^a \hat{n}_j^a-h\sum_i \hat{L}_{i,34}
\label{eq:QRotor}
\end{eqnarray}
The model (\ref{eq:QRotor}) is defined on a coarse-grained lattice, such that each site $i$ contains an even number of Fe atoms.  This is necessary in order to define local operators for the SDW and ODW orders.  The operators $\hat{L}_{i,ab}$ and $\hat{n}_{j}^c$ satisfy the commutation relations (\ref{eq:LLComm}) and (\ref{eq:LnComm}) whenever $i=j$, and they commute for $i\ne j$.  In addition, $\comm{\hat{n}_i^a}{\hat{n}_j^b}=0$ for all $i$ and $j$.  In Eq. (\ref{eq:QRotor}), we allow the couplings $r_a$ to be anisotropic, and also include a Zeeman field that couples to the spin $\hat{S}_z=\hat{L}_{34}$.  Note that higher order terms, which are necessary to bound the energy from below, are not included explicitly in $H_{\rm QR}.$  However,  in a linear spin-wave analysis such terms only lead to weak renormalization of the spectra. The collective mode spectrum is obtained by solving the Heisenberg equations of motion for $\hat{L}_{i,ab}$ and $\hat{n}_i^a$.   Figure \ref{fig:collective} shows the excitation spectrum in the SC and SDW phases. 

Inside the SC, $\Delta\ne 0$, there are three gapped $\zeta$ modes and a gapped $\eta$ mode \cite{podolsky09}.  In addition, the $Q$ mode, shown as a true Goldstone mode of the SC in Fig.~\ref{fig:collective}, is charged and acquires a gap of order the plasma frequency once Coulomb interactions are taken into account.  The $\zeta$ modes can be detected in inelastic neutron scattering experiments, which probe the dynamical spin structure factor
$\chi_{\alpha\alpha}''({\bf k},\omega)=\sum_{m} \l|\langle m | \hat{S}_\alpha(\bf k) | 0\rangle \rr|^2 \delta(\omega-E_{m0}),$
where the sum is over all excited states $|m\rangle$ and $E_{m0}$ is the excited state energy measured from
the ground state.    For the $\zeta_z$ mode, neutron scattering near the wave vector ${\bf Q}$ yields $\chi_{zz}''({\bf k}+{\bf Q},\omega)\propto |\Delta|^2\delta(\omega-\omega_{\zeta_z}({\bf k}+{\bf Q})),$ where $\omega^2_{\zeta_z}({\bf k}+{\bf Q})=|\Delta|^2(r_1-r_5(1-|{\bf k}|^2))/\chi$.  The factor of $|\Delta|^2$ reflects the fact that, although $\zeta_z$ is a particle-particle operator, it can couple to a particle-hole probe such as neutrons inside the SC, where a reservoir of Cooper pairs means that particle number is only conserved modulo 2.  
Similarly, the $\eta$ mode -- which connects the SC and ODW -- can be seen by neutrons with wave vector near ${\bf Q}$, and the amplitude of the signal is also proportional to $|\Delta|^2$. Some of these modes may have been observed in a recent inelastic neutron scattering experiment on
Ba$_{0.6}$K$_{0.4}$Fe$_2$As$_2$ \cite{christianson}.  These modes have also been interpreted as particle-hole bound states in the SC \cite{christianson}.  However, unlike these particle-hole bound states, the energy of the $\eta$ and $\zeta$ modes is independent of temperature, a feature that can distinguish between the two proposals.

Inside the SDW, $\Phi_z\ne 0$, there are two gapless magnon states, and also $\zeta_z$, $\zeta_z^\dagger$, and $R_z$ modes \cite{podolsky09}.  Unlike the superconductor, $\zeta_z^{\dagger}$ does not couple to neutrons in the SDW.  The $R_z$ mode, on the other hand, is a particle-hole spin-one operator.  Therefore, it can naturally be seen in spin-polarized neutron scattering.  For wave vectors near reciprocal lattice vectors ${\bf G}$, $\chi_{zz}''({\bf k}+{\bf G})\propto f({\bf G})\Phi_z^2\delta(\omega-\omega_{R_z}(\bk))$, where $\omega^2_{R_z}({\bf k}+{\bf Q})=|\Phi_z|^2(r_5-r_6(1-|{\bf k}|^2))/\chi$, and $f({\bf G})$ is a form factor that vanishes for ${\bf G}=0$ but is finite for ${\bf G}\ne 0$.  Neutrons near ${\bf Q}$ can also see the $R_z$ mode, with scattering amplitude $\chi_{zz}\propto \Phi_z^2$.

As shown in the lower panel of Fig.~\ref{fig:collective}, the $\zeta$ modes inside the SC are split by an applied Zeeman field, and one of the magnon modes in the SDW becomes gapped under the field.  All other modes are unaffected to linear order in the field.  

{\em Discussion.---}
In the weak coupling RG analysis of Ref. \onlinecite{chubukov}, the SO(6) fixed point (\ref{eq:effInt}) is only reached in the limit of infinitesimal pockets.  In the iron pnictides, the pockets are small but finite.  In this case, a functional RG analysis finds that the interactions eventually flow away from the SO(6) symmetric fixed point \cite{wang}.  However, the RG flows very close to the SO(6) point, which may dominate the finite temperature and energy properties of the system -- the focus of this work.  In some compounds, the size of the pockets may be particularly small, and we predict these to be the best candidates to observe the SO(6) symmetry.

In some iron pnictide materials, such as \loffa, the transition between the SDW and SC phases is accompanied by a structural transition, and is strongly first order \cite{luetkens}.  In these materials, the SO(6) symmetry may be obscured by the structural transition.  Fortunately, other materials such as \coffa\ seem to have continuous transitions between the SC and SDW phases, without a simultaneous structural transition \cite{zhao}.  These materials may provide the most promising candidates for the observation of SO(6) symmetry.  This, in turn, would prove that SC and magnetism have a common origin in the electronic interactions of the iron pnictides.

Finally, we discuss the possibility of observing the ODW state.  Although the ferromagnetic Hund's rule coupling favors SDW order over ODW, other residual interactions could stabilize the ODW in some of the iron pnictides.  Elastic neutron scattering experiments cannot distinguish between SDW and ODW states, both of which give elastic peaks at wave vector ${\bf Q}$.  On the other hand, the ODW does not have low energy magnon states, unlike the SDW.  Thus, inelastic neutron scattering can in principle distinguish between the two states.

{\em Acknowledgements.---} We are very grateful to A. Chubukov and F. Wang for useful discussions. This work was supported by the NSERC, CIFAR, and CRC.

\end{document}